\documentstyle[aas2pp4]{article}

\def\gax    {${_>\atop^{\sim}}$}

\received{27 October 1997}
\accepted{23 April 1988}

\slugcomment{submitted to ApJ Lett, 27 Oct 1997, accepted 23 April 1998}

\lefthead{Garcia et al.}
\righthead{Power Law Decay of GRB Optical Counterparts}

\begin{document}

\title{Power Law Decays in the Optical Counterparts of \\ 
GRB~970228 and GRB~970508}

\author{M. R. Garcia\altaffilmark{1}, 
P. J. Callanan\altaffilmark{2}
D. Moraru\altaffilmark{1}, J. E. McClintock\altaffilmark{1}, 
E. Tollestrup\altaffilmark{1}, S. P. Willner\altaffilmark{1},
C. Hergenrother\altaffilmark{3},
C. R. Robinson\altaffilmark{4}, C. Kouveliotou\altaffilmark{4},
and J. van Paradijs\altaffilmark{5,6}}

\altaffiltext{1}{Center for Astrophysics, 60 Garden St., Cambridge, MA
02138; mgarcia, dmoraru, jmcclintock,etollestrup,swillner, @cfa.harvard.edu}
\altaffiltext{2}{Physics Department, University College, Cork,
Ireland; paulc@ucc.ie}
\altaffiltext{3}{Lunar and Planetary Lab, University of Arizona, Tucson,
AZ; chergen@pirl.lpl.arizona.edu}
\altaffiltext{4}{ Universities Space Research Association and Marshall Space
Flight Center, ES-62, Huntsville, AL 35812;
craig.robinson@msfc.nasa.gov, kouveliotou@eagles.msfc.nasa.gov}
\altaffiltext{5}{University of Amsterdam, Physics Department, 403
Kruislaan, 1098SJ Amsterdam, Netherlands; jvp@astro.uva.nl}
\altaffiltext{6}{Department of Physics, University of Alabama in Huntsville, AL 35899}


\begin{abstract}

We report on $\rm R_c$ and $\rm K$ band observations of the optical
counterpart to GRB~970508 with the Fred Lawrence Whipple Observatory
(FLWO) 1.2m telescope.  Eleven~$\rm R_c$-band observations were
obtained on 1997 May 12, and three~on May~14.  The counterpart clearly
faded between the two nights.  
On May~12 there was no evidence for variability ($<9$\%) on
10-70~minute time scales based on 11 ${\rm R_c}$-band observations. 
On May~19 a one~hour observation set a limit on
the $\rm K$ magnitude of $\rm K >$18.2.  Comparison of these
data points with those obtained by other authors shows that the decay
of the optical counterpart can be well fitted by a power law of the
form $f \sim t^{-\alpha}$ where $\alpha = 1.22 \pm 0.03$ with occasional
fluctuations superposed.  We note that the decay of the optical
counterpart to another burst, GRB~970228, can also be well fitted with
a power law with exponent $\alpha = 1.0^{+0.2}_{-0.5}$ with
occasional fluctuations superposed.  These two decay light curves are
remarkably similar in form to that predicted by cosmic-fireball
models.
\end{abstract}




%

\section{Introduction}

	Since their discovery nearly 30~years ago (Klebesabel, Strong
and Olson 1973), the nature of gamma-ray
bursts (GRBs) has been one of the outstanding problems in
astrophysics.  Bursts with fluxes sufficient to be detected by
CGRO/BATSE are detected approximately once per day, and these bursts
are isotropically distributed on the sky~(\cite{meegan}).  The
observed fluxes of GRBs indicate that this isotropic distribution is
also limited in extent, a fact which has been used to argue that GRBs
are at cosmological distances~(\cite{meegan}).  However, due to the
previous lack of counterparts at other wavelengths, the distances to
GRBs have been uncertain by $\sim 5$ orders of magnitude, leading to
$\sim 10$ orders of magnitude uncertainty in their luminosity.
Progress in understanding GRBs has been hampered by this uncertainty.

The precise locations determined with the Wide Field Camera (WFC) on
board the recently launched Italian-Dutch BeppoSAX observatory have
allowed the discovery of the first optical counterparts to GRBs, for
GRB~970228~(\cite{groot.6584,host.0228}), GRB~970508~(\cite{bond}),
GRB~971214 (\cite{jules}), and GRB~980326 (\cite{iauc6852}).  Three of
these appear to be at cosmological distances: GRB~970228 and
GRB~980326 are surrounded by nebulosity that is most likely a
galaxy~(\cite{host.0228,host.0326}), and GRB~970508 shows optical
absorption lines at redshifts of $z=0.767$ and
$z=0.835$~(\cite{z.85}).  The light curves of the optical counterparts
to GRBs give hints as to the underlying physics of the GRB.  We report
below a modest set of optical and IR photometric observations of the
optical counterpart to GRB~970508.

\section{Observations}

Discovery of the optical counterpart to GRB~970508 was first announced
by Bond~(1997) shortly after BeppoSAX WFC observations yielded an
arcminute location for an X-ray afterglow to the
GRB~(\cite{beppo.970508}).  The counterpart rose to a maximum $\rm R_c
= 19.70 \pm 0.03$ on May 10.77~UT~(\cite{sokolov}), after which it
faded.  Our Cousins R band ($\rm R_c$) observations on May 12.2 UT and
14.2 UT were made during the decay phase, approximately 3.3 and 5.3 days
after the detection of the GRB.  The results in this paper supersede
that reported for the May~12 data in IAUC~6661.

The observations were made with the FLWO 1.2m telescope at
Mt. Hopkins, AZ.  We utilized a $2048\times 2048$ pixel CCD camera
with 0\farcs32 pixels and a standard $\rm R_c$-band filter (the
``Andy-Cam'').  On May~12 conditions were not photometric and the
seeing varied from $2''$ to $3''$; on May~14 conditions were better
but still not photometric.

A journal of the observations and magnitudes is shown in Table~1.
After bias subtracting and flat-fielding, the magnitudes were derived
with DAOPHOT.  Because conditions were not photometric, we have set
our magnitude scale such that the nightly mean magnitude measured for
the star $13''$~N and $4''$~W of the GRB counterpart (star~A in
Table~1) is $\rm R_c =19.49$, as was determined by Sokolov
et~al.~(1997).  The fluctuations in the magnitude of star~A as
measured in each individual exposure reflect both statistical
variations and variable observing conditions.  Included in Table~1 are
the magnitudes of a star with magnitude similar to the GRB, but
presumed to be non-variable (star B in Table~1, $67''$~N and $63''$~W
of the GRB counterpart).  Exposure times were 5~minutes for the first
12~images, and were 20~minutes for the last two images.  Dates are
heliocentric Julian Day at mid-exposure. The mean and standard
deviation ($\sigma$) for each night have been computed directly from
the tabulated magnitudes.

The average magnitudes and mid-exposure times are $\rm R_c =20.23\pm0.02$ and
$\rm R_c =21.03\pm 0.07$, on JD2450580.703 and JD2450582.667,
respectively.  These quoted errors are $= \sigma/\sqrt(N)$, and 
represent the statistical (internal) errors only.  

On May 19.2 UT, approximately 10.3 days after the GRB detection, we
observed the field of GRB~970508 for approximately 1~hour (from HJD
=2450587.7349 to 2450587.7673) with the FLWO 1.2m telescope and the
Smithsonian Astrophysical Observatory 
infrared camera STELIRCAM.  The camera uses an InSb detector array, and the
bandpass was set with a Barr Associates K~filter. 
A total of 45 images, each exposed for 60~seconds, were taken
in a 3x3 grid and then shifted and co-added using standard IR
observing techniques.  Conditions were once again not photometric, and
our magnitude scale has been set by star A, for which  $\rm K =
16.95$~(\cite{aks.pc}).

At the location of the GRB counterpart we find a $1\sigma$ positive
deviation of 28 detected electrons, corresponding to $\rm K = 19.7$.
This does not constitute a detection, so we quote a $4 \sigma$ upper
limit of $\rm K > 18.2$ for the GRB counterpart.  This corresponds to
a flux at $ 2.2 \mu$ of $< 35 \mu$Jy~(\cite{wamsteker,mvz}).

\section{Discussion}

The rms variations evident in the May~12 data 
for GRB~970508 are 6\%, and for the nearby 
 stars A and B are 7\% and 9\%, respectively.  The
difference in the observed rms variations is unlikely to be due to any
intrinsic differences in the objects, but is more likely a statistical
fluctuation due to the modest number of data points (11).
We therefore set a conservative upper limit to any variation
in the GRB counterpart on time scales between 10~and~70~minutes 
of $\leq 9\%$. 

The apparent cosmological distance to GRB~970508 lends credence to the
fireball models for gamma-ray bursts (\cite{goodman,cavallo,Rees}).  In
these models the blast wave accelerates outwards with high Lorentz
factor.  The optical luminosity comes from the interaction of the
blast wave with the surrounding interstellar medium
(\cite{mesz,vietri,sari}).  The GRB was detected  on May
8.904~({\cite{beppo.970508}) so that by May~12 the predicted size
of the  blast wave was $\sim 3$ light-days. 
Thus the fireball model is consistent with our observed lack
of short time scale variability on May~12. 

Some authors have indicated that the optical decay of GRB~970508
(\cite{sokolov}), and also of GRB~970228 (\cite{galama.0228}), was not
well described by a single power-law decay.  In order to test this
possibility, we have fit power-law decay models, $f = a *
t^{-\alpha}$, to the flux densities derived from the magnitudes in
this paper and reported in the literature. For GRB~970508 we use the magnitudes
reported by Sokolov et~al. (1997, 1998), which include measurements
from the 6m SAO~RAS, Keck~(\cite{met.6676}), and Palomar~(\cite{djor})
observatories, transformed to a common $\rm R_c$ bandpass.  To these
we added $\rm R_c$ and R measurements from the
NOT~(\cite{pedersen}), the WIYN, WHT, CAHA, and Loiano
Observatories~(\cite{galama9802160,castro98,schaefer.iauc6658,castro6848}),
from Haute-Provence~(\cite{chev.6663}) and HST~(\cite{fruc.6674}),
corrected (when necessary) to the zero point determined by Sokolov et
al. (1998).  The magnitudes of Castro-Tirado et~al. (1998a)
have been corrected to
the scale of Sokolov et~al. (1998)
by subtracting 0.2 magnitudes
(\cite{goro98}).  These magnitudes are listed in Table~2.  The
corresponding fluxes~(\cite{allen}) are shown in Figure~1.  The fits
do not include the data during the rise of the optical transient (ie,
prior to the maximum on May 10.77), nor after the host galaxy clearly
contributes significantly (after Aug 27).  The best fit $\alpha =
1.19$, but the formal $\chi^2 = 85$ for 43~degrees of freedom.
Clearly a single power-law (alone) is not an adequate description of
the decay.

Motivated by the comments of Fruchter et~al.~1997, we then excluded
all points lying more than $2.5 \sigma$ away from the fit (see Table
2).  The remaining 37~points are well fit ($\chi^2 = 48$) by a single
power law with $\alpha = 1.19\pm 0.02$ (68\% limits, 90\% limits are
$\pm 0.03$).  We note that the slope we find is $2.2\sigma$ and
$0.9\sigma$ different from those found by Sokolov et~al. (1998)
and Galama et~al. (1998)
(respectively), perhaps due to slightly differing
data sets.  The outliers are from data sets that otherwise appear to
fit the curve, which argues that they are not caused by
calibration differences, but are instead real fluctuations in the
decay lightcurve.  This also argues that any difference between
R and ${\rm R_c}$ magnitudes is smaller than the typical error bar.
Galama et~al. (1998) 
measure the magnitude of these fluctuations to be $\sim 15\%$, 
consistent with our findings.

The last two data points included in these fits (from Aug 14.18 and
Aug 26.99) are both $>3\sigma$ above the
power-law fit, indicating that the underlying galaxy may be
contributing significantly to the detected flux.  Recent observations
at KECK (\cite{Bloom1998}) and  WHT (\cite{castro6848}) 
and the SAO (\cite{zharikov.gcn031}) confirm the
existence of a steady component.  A power-law plus constant source 
fit to the data in Table~2 (excluding the same outliers)
finds a decay slope $\alpha = 1.22\pm 0.03$ and a constant source
with with $R_c = 25.6 \pm 0.3$.  This is consistent with the 
magnitude found by 
Zharikov et~al. (1998).

We then repeated the same procedure with the data for GRB~970228 from
Galama et~al. (1997) and Fruchter et~al.  (1997).  Reducing this data
to a common set of $R_c$ magnitudes for the GRB optical counterpart is
complicated by the surrounding nebulosity.  The most recent HST
measurement of this nebulosity finds $V=25.6 \pm 0.25$ (Fruchter
et~al. 1997), so we have corrected the ground based magnitudes for
this refined estimate of the nebular contribution.  We have assumed
that the color of the nebulosity does not change, and therefore the
$V-R = 0.35$ reported by Galama et~al. (1997) indicates $R_{neb} =
25.25 \pm 0.25$.  The measurement of Guarnieri et~al. (1997) took
place in poor seeing, and therefore needs an additional correction due
to contamination by a nearby late type star with $R=22.4 \pm 0.3$. (We
note that this measurement took place on Feb~28.83, not Feb~28.76 as
reported in Galama et~al. 1997.)  The most recent HST measurement of
the optical counterpart finds $V=28.0 \pm 0.25$.  We followed the
method of Galama et~al. in order to convert this V~mag to the $\rm
R_c$ band.  Because the GRB~970228 optical transient became redder
during the decay, we assumed a value of $\rm V-R = 1.0$, redder by 0.1
than the proceeding HST points, and we included the suggested 0.1~mag
uncertainty in the conversion~(\cite{galama.0228}).  For the purposes
of computing $\chi^2$, we treated the upper limit from Mar~04.86 as a
detection one magnitude below the limit, with a one magnitude error.
The fluxes and errors from the literature are listed in
Table~\ref{rflux.228}.

A power law fit to all 11 data points gives $\alpha = 1.04$, but the
resulting $\chi^2 = 19.4$ shows that this fit is an unacceptable
description of the data.  Removing the outlier(s)
from the fit does produce an acceptable $\chi^2$, but unlike
GRB~970508, the results of the fits are dependent upon {\it which}
outlier(s) are removed from the fit.  For example, Fruchter
et~al. (1997) note that the points at Mar~04.86 and Mar~06.32 lie below the
fit, and excluding them produces an acceptable power law fit.  Our
results agree, in that merely removing the point from Mar~06.32
produces an acceptable fit with $\chi^2 = 10.0$, and yields $\alpha =
1.08_{-0.12}^{+0.09}$ (90\% errors).  The point which suffers most
from contamination by surrounding light is that reported by Guarnieri
et~al. (1997),  but it appears to be consistent with this fit. Removing this
point as well results in an insignificant reduction in the scatter to
$\chi^2 = 8.1$ (9 points) and yields $\alpha = 1.11_{-0.12}^{+0.10}$
(90\%).  The single point which has the largest effect on the
scatter is that from Feb~28.99, and removing it alone gives a $\chi^2 =
5.05$ (10 points) and yields $\alpha = 0.75_{-0.21}^{+0.22}$ (90\%).

We conclude that the decay of GRB~970228, like that of GRB~970508, can
be well described by a single power-law, with superposed
fluctuations.  However, perhaps because of the smaller numbers of points
involved, the slope of the decay is not as well determined, and we
conservatively estimate $\alpha = 1.0_{-0.5}^{+0.2}$ (90\%).  
Alternatively, one may choose to describe the decay as two separate
power-laws with different slopes (Masetti et~al. 1997).  We note that
the slope we find is consistent with that found by Masetti et~al. (for
the long term trend) and Fruchter et~al. (1997).  


The spectral slope of the GRB~970508 decay has been measured in the
optical (4000\AA - 6000\AA) to be approximately $F_\nu \sim
\nu^{-1}$~(\cite{z.85,djor}), as predicted in the fireball
models~(\cite{mez.rees}).  Interpolating between the measured $\rm
R_c$ fluxes to the time of our $\rm K$ measurement, this spectral
slope predicts a flux at $2.2\mu$ of $16 \mu$Jy, well below our
measured upper limit of $< 35 \mu$Jy.

In its simplest form, the impulsive cosmological fireball model (eg,
Meszaros and Rees 1997) predicts a single power-law decay.  Given that
the light curves of these two GRB optical counterparts have been
measured for $>100$~days, it is remarkable that, with the
exception of a few fluctuations, they can both be fitted with a single
power law of slope $\alpha = 1.2$.  In the context of the cosmic
fireball models, these fluctuations could be due to inhomogeneities in
the swept-up interstellar medium, or sporadic additional energy input
into the shock front.  
Given the sall number of GRB lightcurve measured, 
we feel it is too early to know whether the power law slope
of $\alpha = 1.2$ is a generic feature of GRB all optical counterparts, or
merely particular to these two.  If it is a generic feature, then it
would argue against a beamed fireball, because beaming produces a wide
variety of power-law decay slopes depending upon the degree and
direction of the beaming (\cite{mesz}).

It therefore may be appropriate to consider spherically symmetric
models for the GRB afterglow.  In these models the power law decay
slope is an indication of either the run of density with radius of the
swept up medium (\cite{mesz,vietri}), or of the shape of the electron
energy distribution (\cite{sari}).

In the spherically symmetric models of \cite{mesz}, the density of the
swept up medium ($\rho$) as a function or radius ($r$) is
parameterized as $\rho \sim r^{-n}$, and the exponent $n$ can be
written as a function of the decay slope and spectral index.  If the
afterglow is radiative, the decay slope of $\alpha = 1.2$ and the
spectral index $F_{\nu} \sim \nu^{-1}$ (as measured in GRB~970508,
\cite{z.85}) imply $\rho \sim r^{-2.7}$ (\cite{mesz}, Eq 5).
However, the fireball is expected to quickly become adiabatic.  Under the
expected conditions of an adiabatic fireball and weak coupling between
the electrons and protons (\cite{mesz}, Eq 8), it is difficult to
produce decay slopes of $\alpha = 1.2$ with density gradients that
might be expected in the ISM.  Typical ISM density gradients would
produce slopes of $\alpha = 1.0$, or slopes steeper than $\alpha =
1.5$, in the spherically symmetric case.

The models of \cite{sari} assume a spherically symmetric fireball
sweeping up an ISM of uniform density, but include the effects of a
power law distribution of electron Lorentz factors $\gamma_e \sim
\gamma_e^{-p} d\gamma_e$.  These models can reproduce the observed
decay slope of $\alpha = 1.2$ for $p = 2.6$, assuming that the optical
frequencies correspond to the 'low frequency' regime, and assuming
that our measurements occur at time $t$ such that $t_m < t < t_c$.
The power law decay should change slope at these critical times, and
the fact that no change is seen implies that
$t_c$\gax 100~days.  We note that the X-ray
decay of some GRBs may be steeper than $\alpha = 1.2$
(\cite{yoshida}).  In the context of the model of \cite{sari}, this
could indicate that the X-rays are in the 'high frequency' regime,
while the optical decay is indicative of the 'low frequency' regime.

We note that AXAF, working in conjunction with satellites designed to
discover GRBs, may be able to provide $\sim 1''$ positions for GRB
counterparts, and also measure both the X-ray spectrum and decay slope to
higher accuracy than has previously been possible.  This should
facilitate 
the search for additional optical and radio counterparts, and should
allow careful testing of models for GRB afterglows.

\acknowledgments

We are grateful to the CfA Telescope TAC for helping to arrange
these GRB observations in an efficient and expedient manner, 
to the anonymous referee for providing several very  helpful
suggestions, and to P. Meszaros and R. Sari for enlightening
discussions.

\begin{table}\label{obs.tab}
\centerline{TABLE~1: Journal of FLWO 1.2m CCD Observations}
\medskip
\begin{center}

\begin{tabular}{cccc}\tableline

  Julian Day	&	 GRB   & B     &  A    \\ \tableline\tableline
  \multicolumn{4}{c}{(May 12.2 UT)}        \\ \tableline
  2450580.6764	& 20.18 & 20.08  &  19.44\\
  2450580.6805	& 20.30 & 20.27  &  19.48\\
  2450580.6851	& 20.34 & 20.30  &  19.43\\
  2450580.6953	& 20.27 & 20.20  &  19.43\\
  2450580.6993	& 20.23 & 20.05  &  19.57\\
  2450580.7032	& 20.17 & 20.15  &  19.57\\
  2450580.7070	& 20.12 & 20.28  &  19.45\\
  2450580.7109	& 20.22 & 20.16  &  19.46\\
  2450580.7148	& 20.23 & 20.13  &  19.59\\
  2450580.7198	& 20.23 & 20.30  &  19.41\\
  2450580.7237	& 20.25 & 20.12  &  19.56\\ \tableline
mean 		& 20.23 & 20.18  &  19.49\\
$\sigma$	& 00.06 & 00.09 &   00.07\\ \tableline
  \multicolumn{4}{c}{(May 14.2 UT)}        \\ \tableline
  2450582.6515 & 21.16 & 20.28 &   19.53\\ 
  2450582.6655 & 20.98 & 20.41 &   19.48\\ 
  2450582.6827 & 20.94 & 20.34 &   19.48 \\  \tableline
mean 		& 21.03 & 20.34 &   19.49 \\ 
$\sigma$	& 00.12 & 00.07 &    00.03 \\ \tableline
\end{tabular}
\end{center}
\end{table}



\begin{table}\label{rflux}
\centerline{TABLE 2:  $\rm R_c,R$ Magnitudes for GRB~970508}
\begin{center}
\begin{tabular}{llcl}\tableline\tableline

Date (UT) 	   &	Magnitude	&Observatory	&Ref \\ \tableline

May 9.128  &	 $21.20\pm	0.1\tablenotemark{a}  $	&CAHA           & 1 \\ 
May 9.195  &	 $21.08\pm	0.15\tablenotemark{a} $	&P200           & 2,3 \\ 
May 9.20   &	 $21.25\pm	0.05\tablenotemark{a} $	&WIYN           & 4 \\ 
May 9.75   &	 $21.19\pm	0.25\tablenotemark{a} $	&SAO            & 5 \\ 
May 9.85   &	 $21.13\pm	0.18\tablenotemark{a} $	&SAO            & 5 \\ 
May 9.899  &	 $20.7 \pm	0.1\tablenotemark{a} $	&CAHA           & 1 \\ 
May 9.93   &	 $20.88\pm	0.05\tablenotemark{a} $	&WHT            & 4 \\ 
May 10.03  &	 $20.46\pm	0.05\tablenotemark{a} $	&WHT            & 4 \\ 
May 10.142 &	 $20.09\pm	0.02\tablenotemark{a} $	&WIYN           & 6,3 \\ 
May 10.178 &	 $19.93\pm	0.09\tablenotemark{a} $	&P200           & 2,3 \\ 
May 10.77  &	 $19.70\pm	0.03 $	&SAO            & 5 \\ 
May 10.850 &	 $19.6 \pm	0.1  $	&LOIANO         & 1 \\ 
May 10.872 &	 $19.6 \pm	0.2  $	&CAHA           & 1 \\ 
May 10.93  &	 $19.80\pm	0.03 $	&SAO            & 5 \\ 
May 10.98  &	 $19.92\pm	0.05 $	&WHT            & 4 \\ 
May 11.01  &	 $19.77\pm	0.07\tablenotemark{b} $	&WHT            & 4 \\ 
May 11.144 &	 $19.9 \pm	0.1  $	&WHT            & 1 \\
May 11.198 &	 $19.87\pm	0.10 $	&P60            & 2,3 \\ 
May 11.76  &	 $20.10\pm	0.03 $	&SAO            & 5 \\ 
May 11.868 &	 $20.2 \pm	0.1  $	&CAHA           & 1 \\
May 12.03  &	 $20.30\pm	0.07\tablenotemark{b} $	&WHT            & 4 \\ 
May 12.135 &	 $20.26\pm	0.03 $	&WIYN           & 6,3 \\ 
May 12.139 &	 $20.3 \pm	0.1  $	&CAHA           & 1 \\
May 12.203 &	 $20.25\pm	0.02 $	&WO             & 7 \\ 
May 12.195 &	 $20.28\pm	0.12 $	&P60            & 2,3 \\ 
May 12.87  &	 $20.63\pm	0.05 $	&SAO            & 5 \\ 
%
%

May 13.179 &	 $20.50\pm	0.15 $	&P200           & 2,3 \\ 
May 13.850 &	 $20.3 \pm	0.1\tablenotemark{b} $	&LOIANO         & 1 \\
May 13.88  &	 $21.09\pm	0.07\tablenotemark{b} $	&SAO            & 5 \\ 
May 14.167 &	 $21.05\pm	0.07 $	&WO             & 7 \\ 
May 14.400 &	 $20.9 \pm	0.2  $	&Haute-Provence & 8,3 \\ 
May 14.860 &	 $21.3 \pm	0.2  $	&LOIANO         & 1 \\
May 14.979 &	 $21.25\pm	0.05 $	&NOT            & 9 \\ 
May 16.884 &	 $21.51\pm	0.10 $	&NOT            & 9 \\ 
May 19.051 &	 $21.88\pm	0.25 $	&NOT            & 9 \\ 
May 19.185 &	 $21.92\pm	0.10 $	&NOT            & 9 \\ 
May 20.875 &     $21.81\pm	0.10 $	&SAO            & 5 \\ 
May 21.892 &	 $22.09\pm	0.07 $	&SAO            & 5 \\ 
May 22.97  &	 $22.04\pm	0.07 $	&WHT            & 4 \\ 
Jun 01.912 &	 $23.10\pm	0.07 $	&NOT            & 9 \\ 
Jun 2.59   &	 $23.1 \pm	0.15 $	&HST            & 10,3 \\ 
Jun 5.26   &	 $23.2 \pm	0.20 $	&KECK           & 11,3 \\ 
Jun 7.879  &	 $23.52\pm	0.10\tablenotemark{b} $	&NOT            & 9 \\ 
Jun 7.917  &	 $23.66\pm	0.10\tablenotemark{b} $	&SAO            & 5 \\ 
Jun 8.991  &	 $23.54\pm	0.20 $	&SAO            & 5 \\ 
Jun 10.928 &	 $23.34\pm	0.20 $	&SAO            & 5 \\ 
Jun 13.966 &	 $23.42\pm	0.14 $	&SAO            & 5 \\ 
Jun 14.9261&	 $23.50\pm	0.25 $	&NOT            & 9 \\ 
Jun 27.893 &	 $23.88\pm	0.16 $	&SAO            & 5 \\ 
July 4.19  &	 $23.95\pm	0.20 $	&WHT            & 4 \\ 
July 7.946 &	 $24.08\pm	0.20 $	&SAO            & 5 \\ 
July 31.843&	 $24.54\pm	0.25 $	&SAO            & 5 \\ 
\end{tabular}
\end{center}
\end{table}

\begin{table}
\centerline{TABLE 2 -- {\it Continued}}
\begin{center}
\begin{tabular}{llcl}\tableline\tableline

Date (UT) 	   &	Magnitude		&Observatory	&Ref \\ \tableline

Aug 2.807  &	 $24.28\pm	0.35 $	&SAO            & 5 \\ 
Aug 14.18  &	 $24.28\pm	0.10\tablenotemark{b} $	&NOT            & 9 \\ 
Aug 26.90  &	 $24.57\pm	0.07\tablenotemark{b} $	&WHT            & 4 \\ 
Oct 9.94  &	$24.30\pm	0.20\tablenotemark{a} $  &SAO		& 12 \\ 
Nov 10.04 &	$24.70\pm	0.15\tablenotemark{a} $	&SAO		& 12 \\ 
Nov 25.97 &	$24.70\pm	0.14\tablenotemark{a} $  &SAO		& 12 \\ 
Nov 29    &	$25.09\pm	0.14\tablenotemark{a} $	&KECK		& 13 \\ 
Jan 24.87 &	$24.96\pm	0.17\tablenotemark{a} $  &SAO		& 12 \\ 
Feb 22.4  &	$25.29\pm	0.16\tablenotemark{a} $	&KECK		& 13 \\ 
Mar 20.5  &	$25.20\pm	0.25\tablenotemark{a} $  &WHT		& 14 \\ 
\tableline
\end{tabular}
\end{center}

\tablenotetext{a}{Data obtained during the rise, or after the 
host galaxy dominates, and excluded from the power-law fit}

\tablenotetext{b}{Outlier dropped from power-law fit}

\tablerefs{
(1) Castro-Tirado et~al. 1998a; \vskip 1pt 	
(2) Djorgovski et~al. 1997; 	
(3) Sokolov et~al. 1997; 	
(4) Galama et~al. 1998; \vskip 1pt		
(5) Sokolov et~al. 1998; 		
(6) Schaefer et~al. 1997; 	
(7) this paper;		\vskip 1pt	
(8) Chevalier and Iloviasky 1997;  
(9) Pedersen et~al. 1997; \vskip 1pt	
(10) Fruchter et~al. 1997; 	
(11) Metzger et~al. 1997; \vskip 1pt	
(12) Zharikov et~al. 1998;	
(13) Bloom et~al. 1998;   \vskip 1pt		
(14) Castro-Tirado et~al. 1998b.	
}

\end{table}

\begin{table}\label{rflux.228}
\centerline{TABLE 3:  $\rm R_c$ Magnitudes for GRB~970228}
\begin{center}
\begin{tabular}{lllc}\tableline

Date(UT)  	& Magnitude & Observatory & Ref \\ \tableline\tableline
%

Feb 28.81	&  $20.5\pm 0.5$				& RAO	&1 \\     
Feb 28.83	&  $21.5^{+0.7}_{-0.5}$\tablenotemark{a}	& BUT    &2 \\
Feb 28.99	&  $20.92\pm 0.15$\tablenotemark{a}		& WHT	 & 1 \\    
Mar 3.10	&  $22.3^{+0.8}_{-0.7}$ 				& APO	& 1 \\    
Mar 4.86	&  $>23.4$					& NOT	& 1 \\
Mar 6.32	&  $24.4^{+0.5}_{-0.4}$\tablenotemark{a}	& KECK	& 1 \\    
Mar 9.90	&  $24.4^{+0.5}_{-0.4}$				& INT	& 1 \\     
Mar 13.00	&  $24.9^{+0.8}_{-0.4}$				& NTT	& 1\\
Mar 26.42	&  $25.17\pm{0.13}$	 			& HST	& 1 \\     
Apr 7.23	&  $25.50\pm{0.13}$				& HST	& 1 \\
Sep 4.7		&  $27.00\pm{0.35}$				& HST	& 3 \\ \tableline
\end{tabular}
\end{center}

\tablenotetext{a}{Outlier excluded from power-law fits}
\tablerefs{
(1) \cite{galama.0228};
(2) \cite{guarnieri.0228}; \vskip 1pt
(3) \cite{iauc6747}
}
\end{table}

%
%
%
%

\clearpage

%
%

\clearpage

\figcaption[apj.final.fig2.ps]{
${\rm R_c}$ band light curves of GRB~970228 (circles, lower line) and
GRB~970508 (boxes, upper line) and best fit power-law decays of
$\alpha = 1.0$ and $\alpha = 1.19$, respectively. Points which have
been excluded from the power-law fits are drawn as open symbols, those
included are drawn as filled symbols.  The curved line is the best fit
to a power law plus constant, and shows that the host galaxy in
GRB~970508 has been detected at a magnitude of ${\rm R_c = 25.6 \pm
0.3}$.  The decays for both GRB are statistically consistent with
power-laws decays with $\alpha = 1.2$ plus occasional small
excursions. \label{lightcurve}}

\clearpage


\begin{thebibliography}{}


\bibitem[Allen 1973]{allen}{Allen, C. W. 1973, Astrophysical
Quantities (London: The Althone Press)} 

\bibitem[Bloom et~al.~1998]{Bloom1998}{Bloom, J. S.,
Kulkarni, S. R.,  Djorgovski, S. G., \&
Frail, D. A. 1998, GCN GRB OBSERVATION REPORT 30}


\bibitem[Bond 1997]{bond}{Bond H. E. 1997, IAUC 6654}

\bibitem[Castro-Tirado et~al. 1998a]{castro98}{Castro-Tirado, A. J.,
et~al. 1998a, Science, 279, 1011}
%

\bibitem[Castro-Tirado et~al. 1998b]{castro6848}{
Castro-Tirado, A. J.,  Gorosabel, J., 
Galama, T., Groot, P., van Paradijs, J., \& 
Kouveliotou, C.  1998b, IAUC 6848}

%

\bibitem[Cavallo and Rees 1978]{cavallo}{Cavallo, G. and Rees, M. J.,
1978 MNRAS 183, 359}

\bibitem[Chary 1997]{aks.pc}{Chary, R. et al. 1998 ApJ 498 L9}

\bibitem[Chevalier and Ilovaisky 1997]{chev.6663}{Chevalier, C. and
Ilovaisky, S. A., 1997, IAUC 6663} 

\bibitem[Costa et al. 1997]{beppo.970508}{Costa, E., et al. 1997, IAUC
6649}
%

\bibitem[Djorgovski et al. 1997]{djor}{Djorgovski, S., et al. 1997,
Nature, 387, 876}
%

\bibitem[Fruchter, Bergeron and Pian 1997]{fruc.6674}{Fruchter, A.,
Bergeron. L., and Pian, E. 1997, IAUC 6674}  

\bibitem[Fruchter et al. 1997]{iauc6747}{Fruchter, A., et~al. 1997,
IAUC 6747}
%

\bibitem[Galama et al. 1997]{galama.0228}{Galama, T. J., et al. 1997
Nature, 387, 479} 

\bibitem[Galama et al. 1998]{galama9802160}{Galama, T. J., et~al. 
1998, ApJ, 497, L13} 	
%

\bibitem[Goodman 1986]{goodman}{Goodman, J., 1986 ApJ 308, L47}

\bibitem[Gorosabel, 1998]{goro98}{Gorosabel, J. U. 1998, private
communication}

\bibitem[Groot et al. 1997a]{groot.6660}{Groot, P. J., et al. 1997a,
IAUC 6660}
%


\bibitem[Groot et al. 1997b]{groot.6584}{Groot, P. J., et al. 1997b IAUC 6584}
%

\bibitem[Groot et~al. 1998]{iauc6852}{Groot, P. J., et al. 1998, 
GCN GRB OBSERVATION REPORT 32}
%
%

\bibitem[Grossan et al. 1998]{host.0326}{Grossan, B., Knop, R.,
Perlmutter, S., \& Hook, I. 1998 GCN GRB OBSERVATION REPORT 35}


\bibitem[Guarnieri et al. 1997]{guarnieri.0228}{Guarnieri, A., et al. 
1997,  A\&A, 328, L13} 
%
%

\bibitem[Halpern et al. 1997]{jules}{Halpern, J.,  Thorstensen,  J.,
Helfand, D.,  Costa, E., \& the BeppoSAX team, 1997, IAUC 6788}

\bibitem[Klebesabel, Strong, and Olson 1973]{grb1}{Klebesabel, R. W.,
Strong, I. B.  \& Olson, R. A.  1973, ApJ, 182, L85}

\bibitem[Masetti et al. 1997]{masetti}{Masetti, N., Bartolini, C.,
Guarnieri, A., \& Piccioni, A. 1997, astro-ph/9711260}

\bibitem[Meegan et al. 1992]{meegan}{Meegan, C.A.,
Fishman, G. J., Wilson, R. B., Paciesas, W. S., Pendleton, G. N., 
Horack, J. M., Brock, M. N., \& Kouveliotou, C. 1992, Nature, 355, 143}

\bibitem[Meszaros and Rees 1997]{mez.rees}{Meszaros, P., and Rees,
M.J. 1997, ApJ, 476, 232} 

\bibitem[Meszaros, Rees, and Wijers 1997]{mesz}{Meszaros, P., Rees,
M. J., \& Wijers, R. A. M. J. 1997, ApJ (subm), astro-ph/9709273}

\bibitem[Metzger, Cohen and Chaffee 1997]{met.6676}{Metzger, M. R.,
Cohen, J. G., \& Chaffee, F. H. 1997, IAUC 6676} 

\bibitem[Metzger et al. 1997]{z.85}{Metzger, M. R., Djorgovski,
S. G., Kulkarni, S. R., Steidel, C. C., Adelberger, K. L., Frail,
D. A., Costa, E., \& Frontera, F. 1997, Nature, 387, 878}

%

\bibitem[Pedersen et al. 1998]{pedersen}{Pedersen, H., et al. 1998,
 ApJ, 496, 311 } 
%
%

\bibitem[Rees and Meszaros 1992]{Rees}{Rees, M. J., \& Meszaros, P.
1992, MNRAS, 258, L41}

\bibitem[Sari, Piran, \& Narayan 1997]{sari}{Sari, R., Piran, T., \& 
Narayan, R. 1997, ApJ, 497, L17} 

\bibitem[Schaefer et al. 1997]{schaefer.iauc6658}{Schaefer, B., et
al. 1997, IAUC 6658}


\bibitem[Sokolov et al. 1997]{sokolov}{Sokolov, V. V.,
Kopylov, A. I., Zharikov, S. V., Costa, E., Feroci, M., Nicastro, L., \& Palazzi, E. 1997,
astro-ph/9709093}  

\bibitem[Sokolov et~al. 1998]{sokolov98}{Sokolov, V. V., Kopylov,
A. I., Zharikov, S. V., Feroci, M., Nicastro, L., \& Palazzi, E. 1998, astro-ph/9802341}

\bibitem[Van Paradijs et al. 1997]{host.0228}{Van Paradijs et al. 1997, Nature, 386, 686}
%

\bibitem[Vietri 1997]{vietri}{Vietri, M. 1997, ApJ, 488, L105}

\bibitem[Wamsteker 1981]{wamsteker}{Wamsteker, W., 1981 A\&A, 97, 329. }

\bibitem[Yoshida et al. 1998]{yoshida}{Yoshida, A., Namiki, M., Otani,
C., Kawai, N., Marakami, T., Ueda, Y., Shibata, R., Uno, S., \& the
ASCA team, 1998, Huntsville GRB Symposium Proceedings,
http://crow.riken.go.jp/~ayoshida/grb.html}  

\bibitem[Zharikov et~al. 1998]{zharikov.gcn031}{Zharikov, S. V.,
Sokolov, V. V., \& Baryshev, Yu. V. 1998, GCN GRB OBSERVATION REPORT 31}

\bibitem[Zombeck 1997]{mvz}{Zombeck, M. 1997, ``Handbook of Space
Astronomy and Astrophysics'', CUP, Cambridge, UK} 

\end{thebibliography}
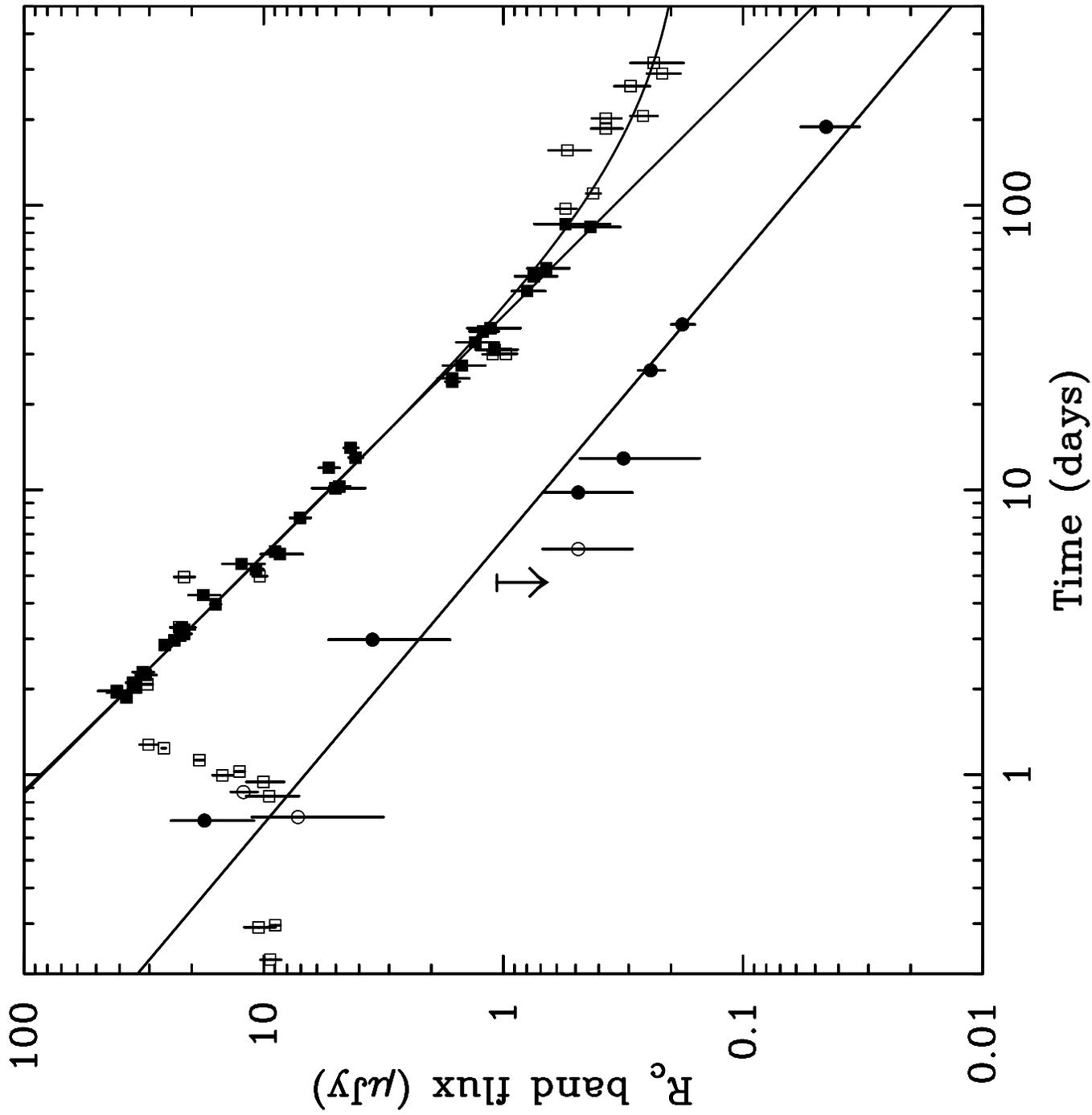
\end{document}